\documentclass[11pt]{article}
\usepackage{latexsym}
\usepackage[latin1]{inputenc}
\usepackage{amsmath,amssymb}
\usepackage{graphicx,ifthen,makeidx,verbatim}
\usepackage{psfrag}
\usepackage{lscape}
\usepackage[dvipsnames]{color,xcolor, colortbl}
\usepackage{arydshln}
\usepackage[babel=true]{csquotes}
\usepackage{bm}
\usepackage{dsfont}
\usepackage[numbers, sort&compress]{natbib}
\usepackage{url}
\usepackage[margin=1.2in]{geometry}
\usepackage{enumerate}

\usepackage{multirow} 
 
\definecolor{Gray}{gray}{0.9}

\newcommand{\mat}[1]{\bm{#1}} 

\newcommand{\argmax}{\mathop{\mathrm{argmax}}}  
\begin{document}

\begin{center}

{\Large \bf Cellwise robust regularized discriminant analysis}
\vspace{1cm}

Stéphanie Aerts$^*$\footnote{\textit{HEC- Liège, University of Liège}, \url{stephanie.aerts@ulg.ac.be}, Phone: +324327272} and Ines Wilms\footnote{\textit{Leuven Statistics Research Centre (LStat), KU Leuven}, \url{ines.wilms@kuleuven.be}}\\
\end{center}

\subsection*{Abstract}
Quadratic and Linear Discriminant Analysis (QDA/LDA) are the most often applied classification rules under normality. In QDA, a separate covariance matrix is estimated for each group. If there are more variables than observations in the groups, the usual estimates are singular and cannot be used anymore. Assuming homoscedasticity, as in LDA, reduces the number of parameters to estimate. This rather strong assumption is however rarely verified in practice. Regularized discriminant techniques that are computable in high-dimension and cover the path between the two extremes QDA and LDA have been proposed in the literature. However, these procedures rely on sample covariance matrices. As such, they become inappropriate in presence of cellwise outliers, a type of outliers that is very likely to occur in high-dimensional datasets.
In this paper, we propose cellwise robust counterparts of these regularized discriminant techniques by inserting cellwise robust covariance matrices. Our methodology results in a family of discriminant methods that (i) are robust against outlying cells, (ii) cover the gap between LDA and QDA and (iii) are computable in high-dimension. The good performance of the new methods is illustrated through simulated and real data examples. As a by-product, visual tools are provided for the detection of outliers.

\subsection*{Keywords}
Cellwise robust precision matrix; Classification;  Discriminant analysis;  Penalized estimation.

\section{Introduction} \label{sec:intro}

Consider a training set $\bm{X} = \{\bm{x}_1,\cdots, \bm{x}_N\}$ of $N$ observations of dimension $p$, each belonging to one of $K$ groups $G_1, \cdots, G_K$, with $n_k$ observations in the $k$-th group and $N = \sum_{k=1}^K n_k$. Discriminant analysis methods aim to construct a decision rule based on $\bm{X}$ that automatically assigns a new observation $\bm{x}$ to one of the $K$ groups. If the group conditional densities  $f_k(\bm{x})$ are known, the Bayes classifier $\delta(.)$ assigns $\bm{x}$ to the group with maximum posterior probability. In \textit{quadratic discriminant analysis} (QDA), where the conditional distributions are assumed Gaussian $N_p(\bm{\mu}_k, \bm{\Sigma}_k)$, this yields the rule
\begin{align}
\delta(\bm{x}) = \mbox{arg}\min_k \left( (\bm{x}-\bm{\mu}_k)^T \mat{\Theta}_k(\bm{x}-\bm{\mu}_k) - \log (\det \mat{\Theta}_k) - 2 \log \pi_k \right),\label{QDArule}
\end{align}
where $\mat{\Theta}_k := \mat{\Sigma}_k^{-1}$ is the $k$-th group precision matrix. This rule splits the measurement space into $K$ disjoint regions with quadratic boundaries. In the special case of \textit{linear discriminant analysis} (LDA), homoscedasticity is further assumed, yielding linear boundaries. Even when the population group precision matrices substantially differ, LDA is often used because it might improve estimation accuracy by reducing the number of parameters to estimate. 

In practice, the group parameters $\bm{\mu}_k$ and $\mat{\Sigma}_k$ are commonly estimated by the arithmetic mean $\bar{\bm{x}}_k$ and the sample covariance matrix $\widehat{\mat{\Sigma}}_k$ in QDA, or the sample pooled covariance matrix $\widehat{\bm{\Sigma}}_{\rm pool}$ in LDA. These procedures will be denoted  by s-QDA and s-LDA from now on. However, when $p \approx n_k$, these estimators become highly inaccurate and for $p > n_k$, their inverse cannot be computed anymore. In the sequel, \textit{high-dimension} refers to such settings. Regularized estimators have been successful in obtaining accurate estimates of $\mat{\Theta}_k$ in high-dimension. They do this by biasing away the estimates from the sample covariance matrices. Particular focus is given to \textit{sparse} precision matrices  where many elements are estimated as zero, see e.g. \cite{YuanWang2013}. \citet{Xu2014} propose to plug the popular \textit{Graphical Lasso} \cite{Friedman2008} sparse precision matrices in quadratic rule (\ref{QDArule}). Similarly, one can obtain a sparse pooled precision matrix for LDA.  However, QDA and LDA are two extreme cases and their underlying assumptions, i.e. all distinct covariance matrices in QDA,  and homoscedasticty in LDA, are rather strong. Therefore, several regularized discriminant methods have been proposed that cover the path between LDA and QDA, see e.g \cite{Friedman2008} or \cite{Price2015}.

All these procedures take the sample covariance matrices $\widehat{\bm\Sigma}_k$ as input. These estimators are however not robust to outliers, i.e. atypical observations. Therefore, the proposed methods inherit their lack of robustness. Since outliers frequently occur in high-dimensional datasets, their possible presence should be accounted for. Several high-dimensional procedures have been proposed to \emph{detect} outliers (see e.g. \cite{Filzmoser2008} or \cite{Zimek2012} for a review). Our focus is on how to \emph{deal} with outliers in regularized discriminant analaysis.  

To robustify the discriminant rule \eqref{QDArule}, one could think of replacing the group means and covariance matrices by standard robust estimates. \citet{CrouxDehon} use the S-estimator, while \citet{HubertVanDriessen} and \citet{Filzmoser} insert the MCD estimator into rule \eqref{QDArule}. Nevertheless, these estimators are not computable anymore in high-dimension. For high-dimensional datasets, robust discriminant methods have been investigated in \cite{HubertEngelen} and \cite{VandenBranden2005}. However, these methods circumvent the high-dimensionality problem by applying a two-step procedure. First, a robust dimension reduction technique is applied. Then a robust discriminant rule, using standard robust location and covariance matrix estimates, is computed in  this low-dimensional subspace.

 Another issue with the standard robust estimators is that they usually downweight an observation even if only one of its components is contaminated. In high-dimensional datasets where many variables are measured on a small number of observations, this may result in a huge loss of information. For such high-dimensional datasets, the  \textit{cellwise contamination} model (see \cite{Alqallaf2009}), where each observation may contain at least one contaminated  component, is more appropriate. The development of cellwise robust procedures only appeared recently, see e.g. \cite{VanAelst2015} and \cite{Agostinelli2015} for low-dimensional datasets, or \cite{Tarr2015} and \cite{Ollerer2015} for high-dimensional ones.

In this paper, we use cellwise robust covariance matrix estimates as an input for regularized discriminant methods. As a result, we obtain discriminant methods that deal with two important topics in applied statistics: regularized estimation and the presence of outliers in high-dimensional datasets. The resulting family of discriminant methods has clear advantages: (i) the methods are robust against cellwise outliers, (ii) as a by-product, they provide a way to detect both rowwise and cellwise outliers,  (iii) they cover the path between LDA and QDA, and (iv) they are computable in high-dimension without requiring an initial dimension reduction technique.

The remainder of this article is structured as follows. In Section \ref{sec:JGL}, we review several non robust regularized discriminant methods. We propose cellwise robust counterparts in Section \ref{sec:robust}. Simulation studies in Section \ref{sec:simul} compare the proposed methods and show their good performance in contaminated and uncontaminated settings. Finally, we analyze two real data sets in Section \ref{sec:real}. We find that the proposed cellwise robust discriminant methods improve the classification performance. Furthermore, two visual tools for outlier detection are provided. The conclusions are outlined in Section \ref{sec:ccl}.

\section{Regularized discriminant methods} \label{sec:JGL}
To classify a new observation $\bm{x}$ in one of the $K$ groups on the basis of the discriminant rule (\ref{QDArule}), we need estimators of the group means $\bm{\mu}_k$ and precision matrices $\bm{\Theta}_k$.  The usual estimators for the group means are the average means $\overline{\bm x}_k$. In this section, we review several procedures to obtain high-dimensional precision matrix estimates that can then simply be plugged into (\ref{QDArule}). Their cellwise robust version will be discussed in Section \ref{sec:robust}.

\paragraph*{GL-LDA and GL-QDA.}
Starting from the sample covariance matrix in the $k$-th group $\widehat{\bm{\Sigma}}_k$,  the \textit{Graphical Lasso} \cite{Friedman2008} maximizes the $L_1$ penalized log-likelihood
\begin{align}
\widehat{\bm{\Theta}}_{k, \rm GL} := \argmax_{\bm{\Theta}_k} \hspace{0.2cm}
n_k \log \det (\bm{\Theta}_k) - n_k \mbox{tr}( \bm{\Theta}_k \widehat{\bm{\Sigma}}_k) -\lambda_1 \sum_{i\neq j} \vert \theta_{k,ij} \vert, \label{GL}
\end{align}
subject to the constraint that $\bm{\Theta}_{k}$ is positive definite. Here, $\theta_{k,ij}$ is the element $(i,j)$ of $\bm{\Theta}_k$ and $\lambda_1  \geq 0$ is a regularization parameter. The $L_1$-norm of the off-diagonal elements ensures that problem (\ref{GL}) can be solved even when the dimension $p$ exceeds the group sample size $n_k$. For large values of $\lambda_1$, many off-diagonal elements in $\widehat{\bm{\Theta}}_k$ will be equal to zero. Under normality, this can be interpreted as conditional independence between the corresponding variables in the specific group. Problem (\ref{GL}) can be solved using the \verb?R?-package  \texttt{huge} \cite{Zhao2012}.  

We denote by GL-QDA, the quadratic classifier obtained by computing the Graphical Lasso in each group and by plugging $\widehat{\bm{\Theta}}_{1, \rm GL} , \dots , \widehat{\bm{\Theta}}_{K, \rm GL}$ into (\ref{QDArule}), see \cite{Xu2014}. A regularized estimate of the pooled precision matrix can be obtained in a similar way by using 
\begin{align}
\widehat{\bm{\Sigma}}_{\rm pool}:= \sum_{k=1}^K \dfrac{n}{n-K} \widehat{\mat{\Sigma}}_k
\end{align}
 as input in (\ref{GL}). We denote by GL-LDA the resulting linear classifier.

\paragraph*{JGL-DA.}
The GL-QDA discriminant rule does not exploit the potential similarities between the groups since it estimates the $K$ precision matrices independently. On the other hand, the homoscedasticity assumption behind GL-LDA ignores the group specificities that may be of particular interest in the classification context. To encourage similar sparsity patterns across the groups, \citet{Price2015} (and \citet{Gao2016} in model based clustering) use the \textit{Joint Graphical Lasso} (JGL)  \cite{Danaher2014}. The JGL estimates are
\begin{align}
\left(\widehat{\bm{\Theta}}_{k, \rm JGL}\right)_{k=1}^K := \argmax_{\bm{\Theta}_1,\dots , \bm{\Theta}_K}\hspace{0.2cm}
 \sum_{k=1}^{K} n_k \log &\det (\bm{\Theta}_k) - n_k \mbox{tr}( \bm{\Theta}_k \bm{\widehat{\Sigma}}_k)  \nonumber \\
 &- \lambda_1\sum_{k=1}^{K} \sum_{i\neq j} \vert \theta_{k,ij} \vert
 - \lambda_2 \sum_{k<k'} \sum_{i,j} \vert \theta_{k,ij} - \theta_{k',ij}\vert, \label{JGL}
\end{align}
subject to the constraint that $\bm{\Theta}_1, \dots, \bm{\Theta}_K$ are positive definite. The first penalty in \eqref{JGL}, with regularization parameter $\lambda_1 \geq 0$, is the same as in \eqref{GL}. The second penalty in (\ref{JGL}), with regularization parameter $\lambda_2 \geq 0$, encourages similar sparsity patterns across the groups and similar signs and values for the non-zero elements. The GL-QDA estimator corresponds to the particular case with  $\lambda_2=0$.  Large values of the similarity parameter $\lambda_2$ yield precision matrices with many similar elements across the groups.  Varying the parameter $\lambda_2$ provides a variety of classifiers that lie in between LDA and QDA. Problem (\ref{JGL}) can be solved using the \texttt{R}-package \texttt{JGL} \cite{JGL}.  We denote by JGL-DA the discriminant rule obtained by plugging the precision matrices of equation (\ref{JGL}) into rule (\ref{QDArule}).

\paragraph*{RDA.}

\citet{Friedman1989} proposes another regularized discriminant method, denoted by RDA from now on. Like JGL-DA, it gives a path from LDA to QDA and it is computable in high-dimension. Unlike JGL-DA, it does not provide sparse precision matrix estimates.

RDA starts by computing  a convex combination of the group specific and pooled sample covariance matrices
\begin{align}
\bm{\widehat{\Sigma}}_k^{\rho_1}= (1-\rho_1)\bm{\widehat{\Sigma}}_{k} + \rho_1  \bm{\widehat{\Sigma}}_{\rm pool}, \label{RDA1}
\end{align}
where $0 \leq \rho_1 \leq 1$ is a regularization parameter.
Then, the resulting estimator (\ref{RDA1}) is shrunken towards a multiple of the identity matrix 
\begin{align}
\bm{\widehat{\Sigma}}_{k,\rm RDA} = (1- \rho_2) \bm{\widehat{\Sigma}}_k^{\rho_1}   +  \frac{\rho_2}{p} \mbox{tr}(\bm{\widehat{\Sigma}}_k^{\rho_1}) \bm{I}_p, \label{RDA2}
\end{align}
with a second regularization parameter $ 0 \leq \rho_2 \leq 1$.
The s-QDA solution corresponds to $\rho_1 = \rho_2=0$ while the s-LDA solution is obtained with $\rho_2=0,\rho_1=1$. The resulting estimators $\bm{\widehat{\Sigma}}_{k,\rm RDA}$  can then  be inverted to obtain regularized precision matrix estimates to be plugged into (\ref{QDArule}).

\section{Cellwise robust discriminant methods}
\label{sec:robust}

The estimators from Section \ref{sec:JGL} all use the sample covariance matrices and/or the pooled covariance as input and are therefore not robust against cellwise outliers. To obtain cellwise robust discriminant methods, we start with computing initial cellwise robust covariance matrices $\bm{S}_k$ and the corresponding pooled covariance $\bm{S}_{\rm pool}$. These cellwise robust covariance matrices are used to replace the sample covariance matrices $\widehat{\bm \Sigma}_k$ and $\widehat{\bm \Sigma}_{\rm pool}$ as input of  the Graphical Lasso in (\ref{GL}), the Joint Graphical Lasso in (\ref{JGL}) or RDA in (\ref{RDA1}) and (\ref{RDA2}).  Then, we use these precision matrices along with robust mean estimates, namely the vector of marginal medians, in the discriminant rule (\ref{QDArule}). As a result, we obtain discriminant methods that are both robust against cellwise outliers and easily computable in high-dimension. 

These cellwise robust counterparts of the regularized discriminant methods of Section \ref{sec:JGL} are denoted by rGL-LDA, rGL-QDA, rJGL-DA and rRDA from now on. The code to obtain these estimators is made available on \url{http://feb.kuleuven.be/ines.wilms/software}.

\subsection{Cellwise robust covariance matrix estimates}
\label{CellwCov}

 We estimate each bivariate covariance between variables $X^i$ and $X^j$ by
\begin{align}
s_{ij} = \widehat{\mbox{scale}}(X^i)\widehat{\mbox{scale}}(X^j) \widehat{\mbox{corr}} (X^i,X^j), \label{pairw}
\end{align}
as in \citet{Ollerer2015}. We use the robust $Q_n$-estimator \cite{RousseeuwCroux1993} as $\widehat{\mbox{scale}}(.)$   and the Kendall's correlation estimator as $\widehat{\mbox{corr}}(.)$ . For a bivariate sample ${(x_1^i,x_1^j), \dots , (x_n^i,x_n^j)}$, this correlation estimator  is defined as 
\begin{align*}
\widehat{\mbox{corr}}_{\rm K}(X^i, X^j) = \frac{2}{n(n-1)}\sum_{l<m} \mbox{sign}\left( (x^i_l-x^i_m)(x^j_l-x^j_m)\right).
\end{align*}
By using signs rather than numerical values, Kendall correlation is a robust correlation measure and, hence, can cope with outliers.

 A $\mathcal{O}(n\log n)$ algorithm to compute it is available in the \texttt{pcaPP} package \cite{pcaPP}. 
The covariance matrices obtained by estimating each pairwise covariance  as in (\ref{pairw}) are denoted by $\bm{S}_k$ and the corresponding pooled covariance by $\bm{S}_{\rm pool}$. \citet{Ollerer2015} show that replacing the sample covariances by these robust estimates in the Graphical Lasso estimator (\ref{GL}) results in robust precision matrices with 50\% breakdown point against cellwise contamination. Note that few proposals of cellwise robust covariance estimators that are computable in high-dimension are available in the literature. Alternatives can be found in \cite{Alqallaf2002} and \cite{Tarr2015}. 

\subsection{Selection of the regularization parameters}

The regularized methods rGL-LDA, rGL-QDA, rJGL-DA and rRDA all depend on one or two regularization parameters. To select the regularization parameters of a given method, we consider a grid of values, i.e. a one-dimensional grid for rGL-QDA and rGL-LDA, and a two-dimensional grid for rJGL-DA and rRDA.  For each (combination of) regularization parameter(s), we compute the corresponding precision matrix estimates $\widehat{\bm\Theta}_1,\dots ,\widehat{\bm\Theta}_K$ (or  $\widehat{\bm \Theta}_{\rm pool}$ for LDA) and we search for the optimal ones minimizing the  Bayesian Information Criterion
\begin{align}
\mbox{BIC} = \sum_{k=1}^K \left[ n_k \mbox{tr} \left(\bm{S}_k \widehat{\mat{\Theta}}_{k}\right) - n_k \log (\det( \widehat{\mat{\Theta}}_{k}))\right] + \log(N) \mbox{df},\label{BIC}
\end{align}
where  $\mbox{df}$ denotes the degrees of freedom of the model. The cellwise robust covariance matrices $\bm{S}_k$ are replaced by $\bm{S}_{\rm pool}$ in  LDA. Note  that $\widehat{\bm\Theta}_1,\dots ,\widehat{\bm\Theta}_K$ and $\mbox{df}$ depend on the regularization parameters. 

We take the degrees of freedom $\mbox{df}$ to be the total number of \textit{distinct} non-zero elements in $ \widehat{\mat{\Theta}}_{1}, \dots, \widehat{\mat{\Theta}}_{K}$. \citet{Danaher2014} replace $\mbox{df}$ by the total number of non-zero elements in the estimated precision matrices. As such, model complexity is however overestimated since it does not take into account the fact that some elements may be identical across the $\widehat{\bm{\Theta}}_k$.

\paragraph*{Grid bounds.} For each regularization parameter, we consider a logarithmic spaced grid of five values between  chosen upper and lower bounds, i.e. the grid consists of the exponential of five equally spaced values between the logarithms of the bounds. The lower bound is a tenth of the upper bound. Below, we discuss the choice of the upper bound for each of the proposed methods. Up to our knowledge, no upper or lower bound for such a grid is proposed in the literature for the multiple group setting.

For rGL-QDA, we take as upper bound for $\lambda_1$ 
\begin{align}
\lambda_{1,\rm max}^{\rm rGL-QDA}= \max_k \max_{i,j} n_k \vert (\bm{S}_k)_{ij} - \bm{I}_{ij}\vert, \label{upGL}
\end{align}
 where $\bm{I}$ is the identity matrix. 
This value is the maximum over the $K$ upper bounds considered by default in the \texttt{huge} package when performing the Graphical Lasso in each group. The upper bound for rGL-LDA can be obtained by replacing $\bm{S}_k$ by $\bm{S}_{\rm pool}$ in (\ref{upGL}). 

For rJGL-DA, we consider 
\begin{align*}
\lambda_{1,\rm max}^{\rm rJGL-DA}= \max_k \max_{i,j ; i \neq j} n_k \vert (\bm{S}_k)_{ij} \vert,
\end{align*} as upper bound for $\lambda_1$ since $\lambda_1 \geq \lambda_{1,\rm max}$ is a sufficient condition for all the off-diagonal elements of the solution to be zero \cite{Danaher2014}. For $\lambda_2$, we propose the heuristic upper bound 
\begin{align*}\lambda_{2,\rm max}^{\rm rJGL-DA} = \max_{k}  \max_{i,j} n_k \vert (\bm{S}_{\rm pool})_{ij} - (\bm{S}_k)_{ij} \vert.
\end{align*}
We take this bound since $\lambda_2 > \lambda_{2,\rm max}^{\rm rJGL-DA}$ is a necessary condition for $\bm{S}_{\rm pool}$ to satisfy the KKT conditions of problem \eqref{JGL} (for $\lambda_1=0$ , if $\bm{S}_{\rm pool}$ has full rank and $K=2$). 

For rRDA, the regularization parameters  $\rho_1$ and $\rho_2$  in equations \eqref{RDA1} and \eqref{RDA2} are the coefficients of convex combinations. Hence, we set the upper bounds to one.

\section{Simulation study}  \label{sec:simul}

In this section, we compare the performance of the regularized discriminant methods (cfr. Section \ref{sec:JGL}) and their cellwise robust counterparts (cfr. Section \ref{sec:robust}) through a simulation study. The considered non robust methods are s-LDA, s-QDA, GL-LDA, GL-QDA, JGL-DA and RDA. The cellwise robust counterparts are respectively r-LDA, r-QDA, rGL-LDA, rGL-QDA, rJGL-DA and rRDA. The former two are obtained by plugging the cellwise robust covariance matrices $\bm{S}_k$ (or the pooled matrix $\bm{S}_{\rm pool}$) directly into rule \eqref{QDArule}. 
Regularization parameters for the regularized methods are selected based on the BIC equation \eqref{BIC}. For the non robust versions, we use $\widehat{\bm \Sigma}_k$ instead of $\bm{S}_k$ in the BIC equation \eqref{BIC}. 

Up to our knowledge, no comparison of all the considered  discriminant methods has been made in the literature. Hence, it is worth analyzing their relative performances in depth. To this end, we consider both uncontaminated and contaminated settings.

\paragraph*{Performance measures.} 
For all the settings described below, we simulate 1000 training datasets consisting of $K$ groups for each of which we estimate the mean and precision matrix. These  estimates are then used to construct a discriminant rule. Additionally, for each training dataset, we generate a test dataset that consists of $N_{\rm test}=1000$ observations. For each observation  of the test set, we randomly select (with equal probability) one of the $K$ populations, then randomly draw a value from this population. We evaluate the discriminant methods in terms of \textit{classification performance} and \textit{estimation accuracy}.

To evaluate classification performance, we use the test datasets to compute the \textit{average percentage of correct classification} over the 1000 simulation runs. The higher the average percentage, the better the classification performance. 

To measure estimation accuracy, we report the \textit{Kullback Leibler (KL) distance}. Under the normal model, the KL-distance from the model with estimated precision matrices  $\widehat{\bm{\Theta}}_1,\dots ,\widehat{\bm{\Theta}}_K$  to the model with true precision matrices $\bm{\Theta}_1,\dots ,\bm{\Theta}_K$ is
\begin{align*}
\mbox{KL}(\widehat{\bm{\Theta}}_1,\dots ,\widehat{\bm{\Theta}}_K ;\bm{\Theta_1},\dots, \bm{\Theta_K})= \left(\sum_{k=1}^{K} - \log \det(\widehat{\bm{\Theta}}_k \bm{\Theta}_k^{-1}) + \mbox{tr}(\widehat{\bm{\Theta}}_k \bm{\Theta}_k^{-1}) \right)- Kp.
\end{align*} The lower the KL-distance, the more accurate the estimates. If all the estimates are equal to the true precision matrices, the KL-distance is equal to zero.

\subsection{Uncontaminated scheme} \label{sec:uncontam}
We consider two different scenarios. In the first one, the number of groups is set to $K=10$, with group sample sizes $n_k=30$ and varying dimension $p=5, 10, 30$. The precision matrices of the first five groups are equal. They have diagonal elements equal to one and zero off-diagonal elements except in the upper left block of size $2 \times 2$, where the off-diagonal elements are set to 0.9. For the precision matrices of the next five groups, the $2 \times 2$ block with 0.9 off-diagonal elements is located in the lower right corner. The mean vectors of the first five groups have elements all equal to zero except element $k$ in the $k$-th group that is equal to 3. The mean vectors of the remaining five groups have the opposite sign.

In the second scenario, the number of groups is set to $K=6$, with group sample size $n_k=30$ and dimension $p=50$. The covariance matrices are chosen as in \cite{Friedman1989}: they  are all diagonal with elements $\left( 9(i-1)/(p-1)+1\right)^2$ for $i=1,\dots , p$ in the first three groups, and $\left( 9(p-i)/(p-1)+1\right)^2$ for $i=1,\dots , p$ in the three other groups. The condition number is thus the same for all the groups but the low and high variance subspaces of groups 1 to 3 and 4 to 6 are complementary. All the elements of the mean vector of group $k$ are set to zero except the $k$-th element in the first three groups, and the $p-k$-th element in the three other groups that are equal to $\log(p)$.

\paragraph*{Results.}

Table \ref{TabSimul1} gives the correct classification percentages and KL-distances for the non robust (top) and robust methods (bottom) in the two scenarios. Robust methods are generally expected to perform well also in uncontaminated settings. In both scenarios, the correct classification percentages and KL-distances of the robust techniques are, overall, quite similar to those of the non robust ones, regardless of the dimension $p$. The use of the cellwise robust discriminant methods (instead of the non robust ones) thus only results in a small statistical efficiency loss.

Next, we compare the standard discriminant methods s-LDA and s-QDA to the regularized ones GL-LDA, GL-QDA. For their cellwise robust counterparts, the same conclusions can be made. In a low dimension (Scenario 1, $p=5$), all the methods show similar classification performance. As soon as the dimension increases (Scenario 1, $p=10$ and $p=30$), GL-LDA and GL-QDA perform much better.  Regularization is  necessary to improve both the classification performance and estimation accuracy of the standard methods. For instance, in dimension $p=10$, the standard quadratic classifier s-QDA  suffers from low estimation accuracy, i.e. its KL-distance is 3 times that of its regularized version GL-QDA. This, in turn, negatively impacts its classification performance. In dimension $p=30$, s-QDA is not computable anymore as indicated by \texttt{NA} in Table \ref{TabSimul1}. Its regularized version, on the contrary, is always computable and yields good classification performance and high estimation accuracy. Also for the robust methods, although still computable in this setting\footnote{Kendall's coefficient uses all the possible pairs of observations. As such, the corresponding covariance matrices may be invertible for $p> n_k$ but they become singular as soon $p> n_k(n_k-1)/2$. }, r-QDA  yields poor classification performance and estimation accuracy compared to rGL-QDA.

Further improvement over GL-LDA and GL-QDA can be obtained by using JGL-DA. The latter not only yields a high correct classification percentage but also the most accurate precision matrix estimates in each setting. When several groups share similar sparsity patterns (as in the considered simulation settings), estimation accuracy can be considerably improved by using a method that covers the path between LDA and QDA, like JGL-DA.  In other (unreported) simulation studies expected to favour  either GL-LDA or GL-QDA, we also find JGL-DA to be a tough competitor with respect to both classification performance and estimation accuracy.  The results are available from the authors upon request. The same overall conclusions hold when comparing the cellwise robust discriminant methods.

Differences between the regularized methods are even more marked in the second scenario (see Table \ref{TabSimul1}).  JGL-DA still attains the best correct classification performance and estimation accuracy, followed by GL-QDA. The standard s-QDA is not computable since $p$ is too large. The methods GL-LDA and RDA show poor classification performance and low estimation accuracy. In this scenario, the groups are more difficult to distinguish since they have complementary low and high variance subspaces, and the mean differences lie in the direction with the lowest variance. The precision matrix estimates used in GL-LDA, RDA (and s-LDA) artificially inflate the lowest variances by pooling and enlarging the smallest eigenvalues.  This results in poor performance of these techniques. JGL-DA, on the contrary, lies in between LDA and QDA without shrinking the precision matrices towards each other as a whole. Only the coefficients that are similar across groups will be estimated identically. Hence, JGL-DA shows good performance in terms of both correct classification and estimation accuracy.

\begin{table}[t]
\begin{footnotesize}
\begin{center}
\caption{Percentages of correct classification (CC) and KL-distance for the non robust discriminant methods (top) and their cellwise robust counterparts (bottom), averaged over 1000 simulation runs.
}\label{TabSimul1}
\begin{tabular}{l|ll|cccccc}
\cline{4-9}
\multicolumn{3}{c|}{}& s-LDA & s-QDA & GL-LDA & GL-QDA & JGL-DA &RDA  \\
\hline
\multirow{6}{*}{\parbox{1.5cm}{Scenario 1  $K=10$ }}  &$p=5$	&CC&  78.4 & 79.0 & 78.5 & 81.4& 81.9 & 78.4   \\ 
&& KL & 12.65 & 7.53  & 12.73 & 3.60 & 3.01 & 12.64 \\ 
\cline{2-9}
&$p=10$	&CC & 82.7 & 76.6  & 83.4 & 85.6 & 86.1& 82.8 \\ 
& &KL & 14.06 & 39.36  & 13.46 & 13.60 & 3.68& 14.00 \\ 
\cline{2-9}
& $p=30$	&CC & 77.7 &  \texttt{NA} & 80.5 & 83.0& 83.5 & 75.4  \\ 
&&KL &  30.29& \texttt{NA}  & 21.87 & 40.41& 5.03 & 58.67  \\ 
   \hline  
\multirow{2}{*}{\parbox{1.5cm}{Scenario 2 $K=6$}} &$p=50$	 & CC & 23.5 & \texttt{NA} & 25.7 & 54.5 & 71.4 & 25.7 \\ 
&& KL&   223.90 & \texttt{NA} & 158.12 & 85.43 & 78.67 & 155.72  \\ 
   \hline  
  \multicolumn{9}{c}{}\\
\cline{4-9}
\multicolumn{3}{c|}{}& r-LDA & r-QDA& rGL-LDA & rGL-QDA  & rJGL-DA &rRDA\\
\hline
\multirow{6}{*}{\parbox{1.5cm}{Scenario 1  $K=10$ }} 
& $p=5$ &CC  & 77.0 & 78.2  & 76.9 & 79.3 & 79.6 & 77.0\\ 
&& KL & 14.96 & 8.16  & 15.97 & 8.90 & 8.15& 15.07  \\ 
\cline{2-9} 
&$p=10$ &CC  & 81.4 & 69.3  & 82.2 & 84.5 & 85.1 & 81.4 \\ 
&& KL &  14.57 & 104.11  & 14.00 & 13.88 &  4.44 &14.51\\ 
\cline{2-9}
&$p=30$ & CC & 76.1 & 59.7  & 77.4& 79.7 & 80.1& 73.5 \\ 
&& KL & 22.86 & 139.18  & 22.98 & 44.57& 11.02 & 59.01  \\ 
   \hline
\multirow{2}{*}{\parbox{1.5cm}{Scenario 2 $K=6$}} 
& $p=50$ &CC & 24.0 & 70.2  & 24.6 & 51.9& 61.4 &  25.5 \\
&& KL & 177.28 & 238.08  & 159.64 & 93.01  & 104.73& 156.07
 \\
 \hline
   \end{tabular}
    \end{center}
    \end{footnotesize}

\end{table}

\subsection{Contaminated scheme}
We compare the performance of the non robust and robust discriminant methods in the presence of cellwise outliers. To this end, we add contamination to the settings from Section \ref{sec:uncontam}.  In each training set, we randomly replace a given proportion of the cells in each group. The considered cellwise contamination percentages are $\varepsilon=5\%$ and 10\% in the first scenario and 1\% in the second one. The test datasets are generated as in Section \ref{sec:uncontam} without contamination. 

In the first scenario, each contaminated cell is drawn from a normal distribution $N(-10, 0.2)$ in the first five groups and from $N(10,0.2)$ in the five other groups. This shift contamination may drive the estimated means of the first five groups in the direction of the means of the remaining groups (and vice-versa) and inflate the sample covariance estimates. In the second scenario, each contaminated cell is drawn from a normal distribution with a large variance $N(0,50)$.

\begin{figure}
\begin{minipage}[t]{.46\linewidth}
\text{5\% cellwise contamination }

$p=5$

\includegraphics[scale=0.33]{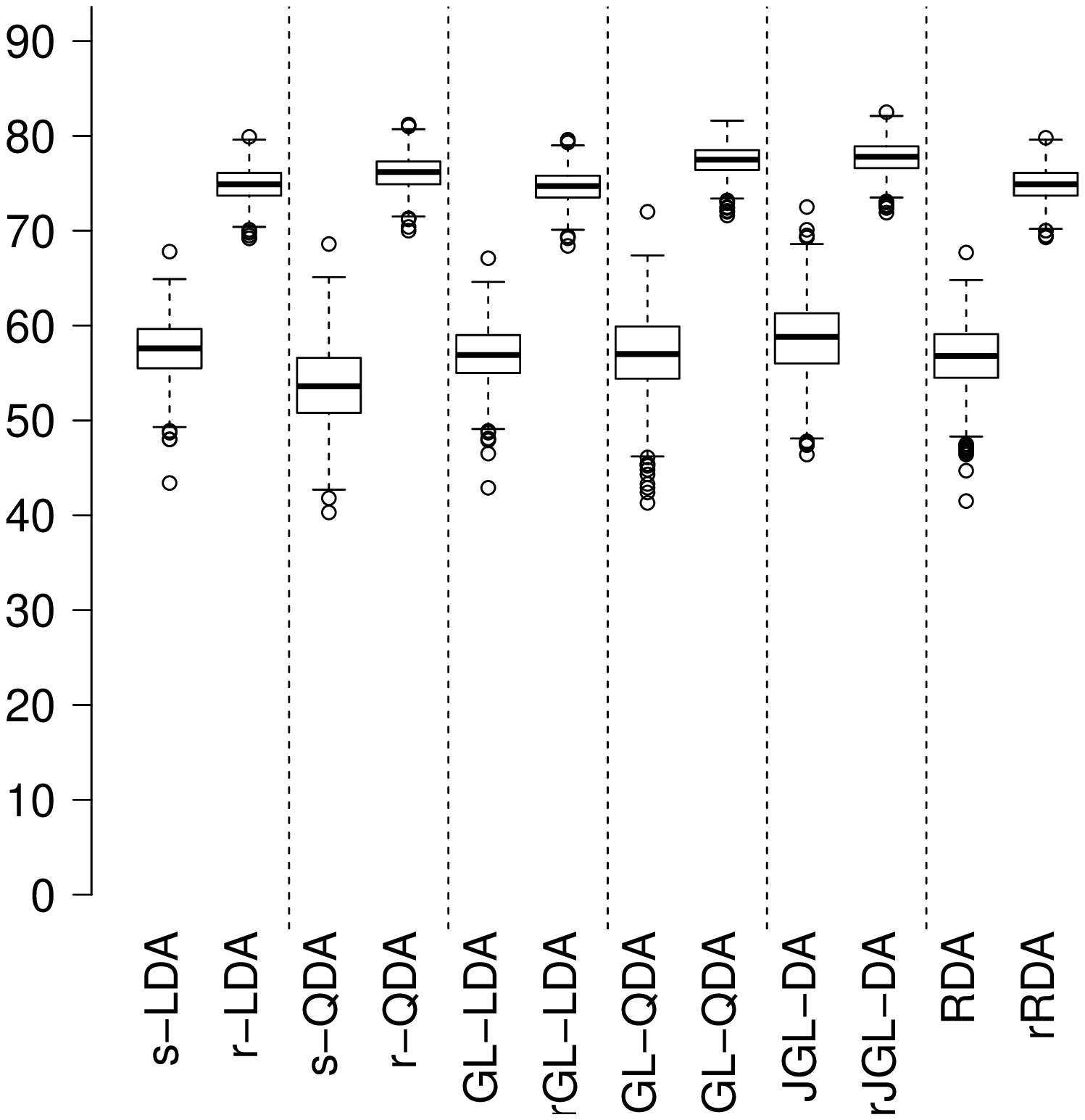}

$p=10$

\includegraphics[scale=0.33]{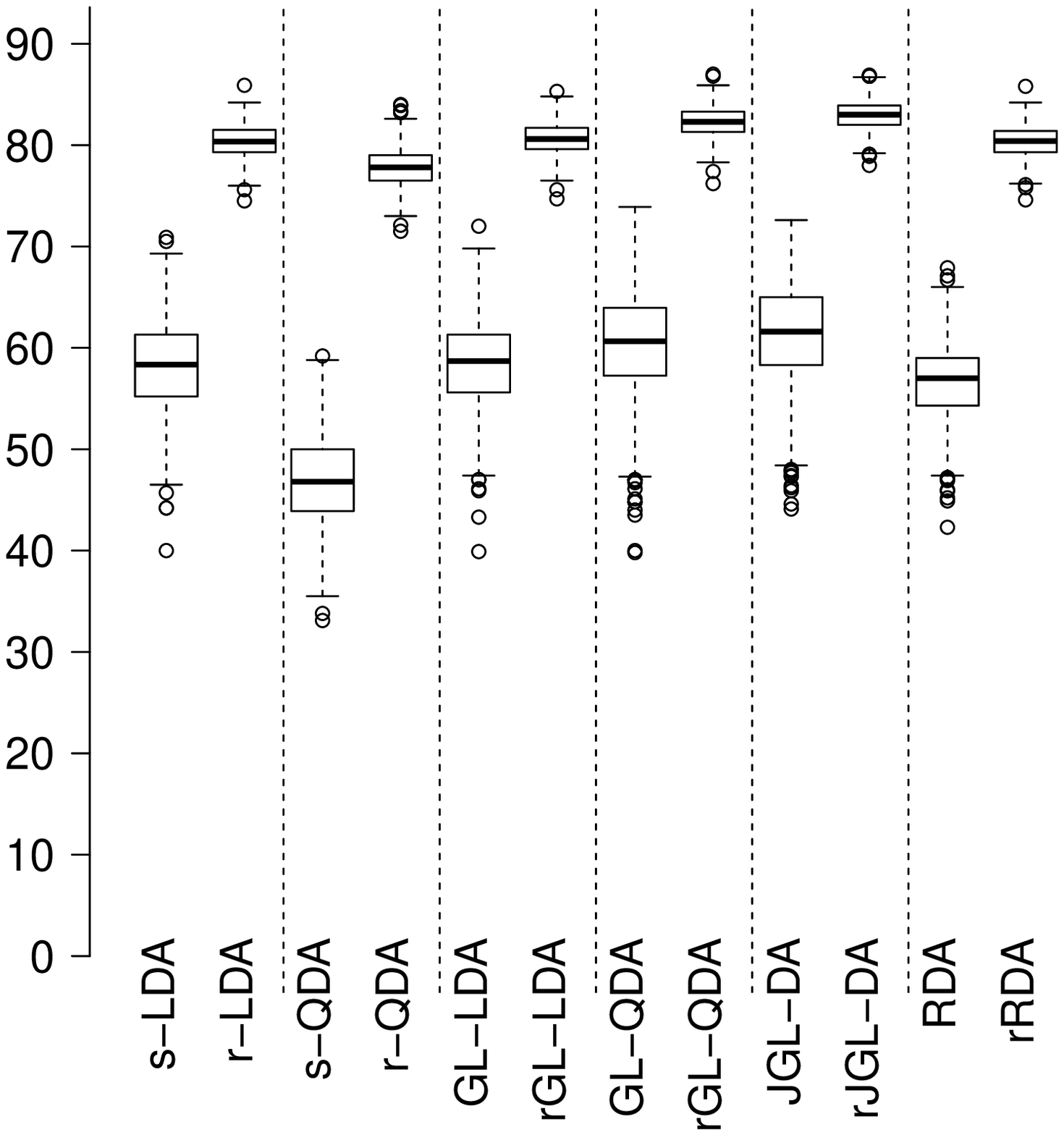}

$p=30$

\includegraphics[scale=0.33]{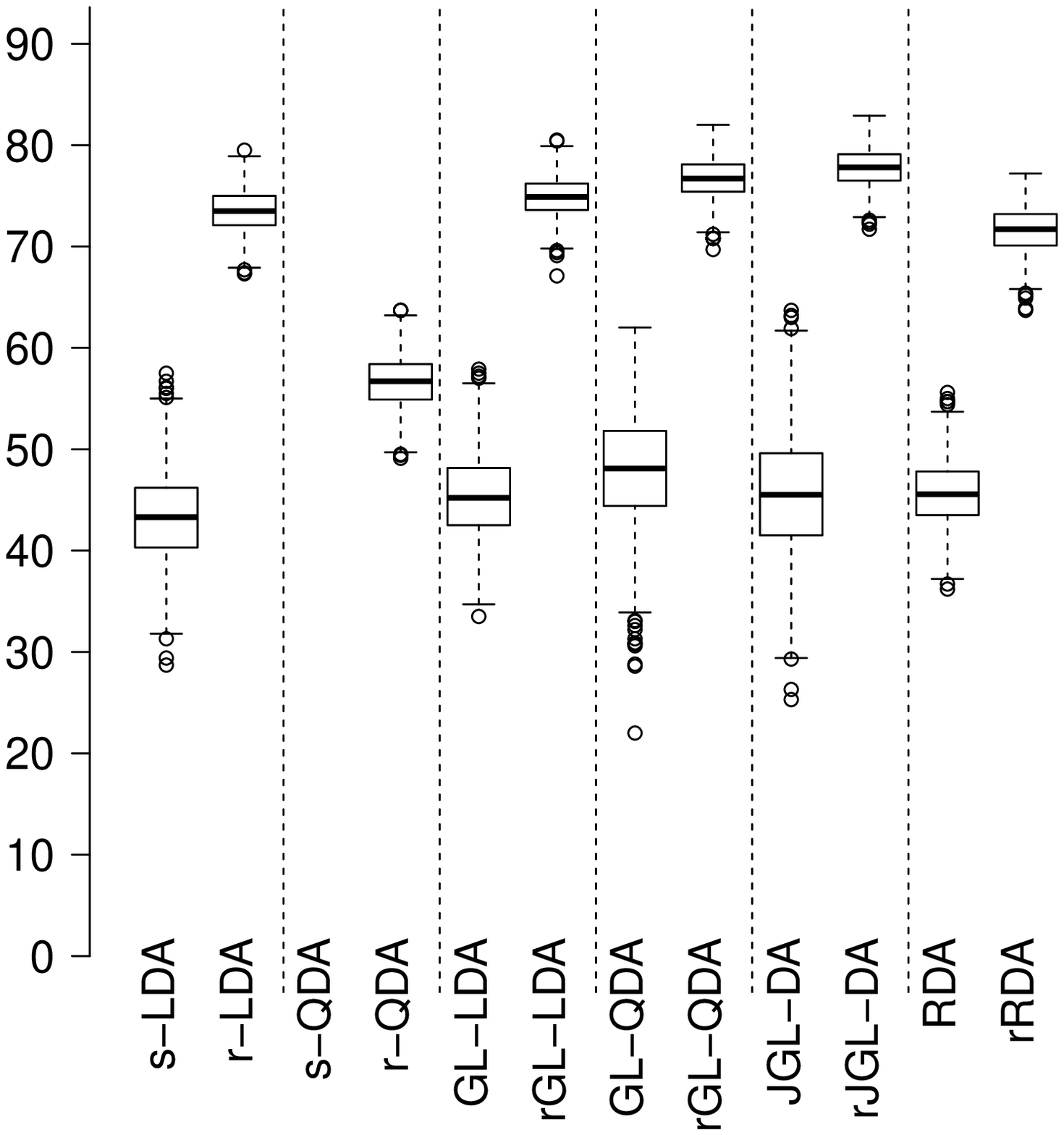}
\end{minipage}\hfill
\begin{minipage}[t]{.46\linewidth}
\text{10\% cellwise contamination}

$p=5$

\includegraphics[scale=0.33]{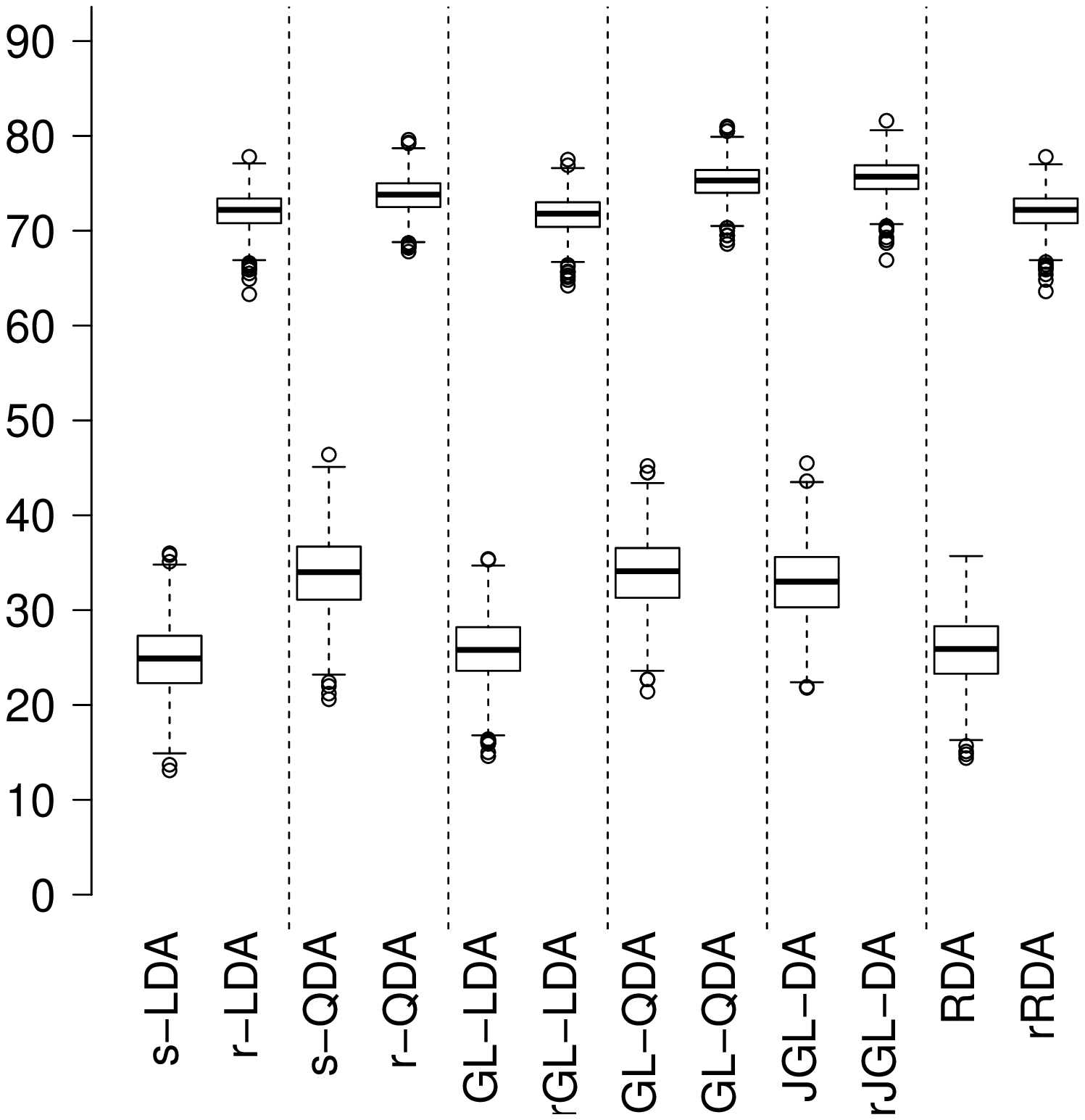}

$p=10$

\includegraphics[scale=0.33]{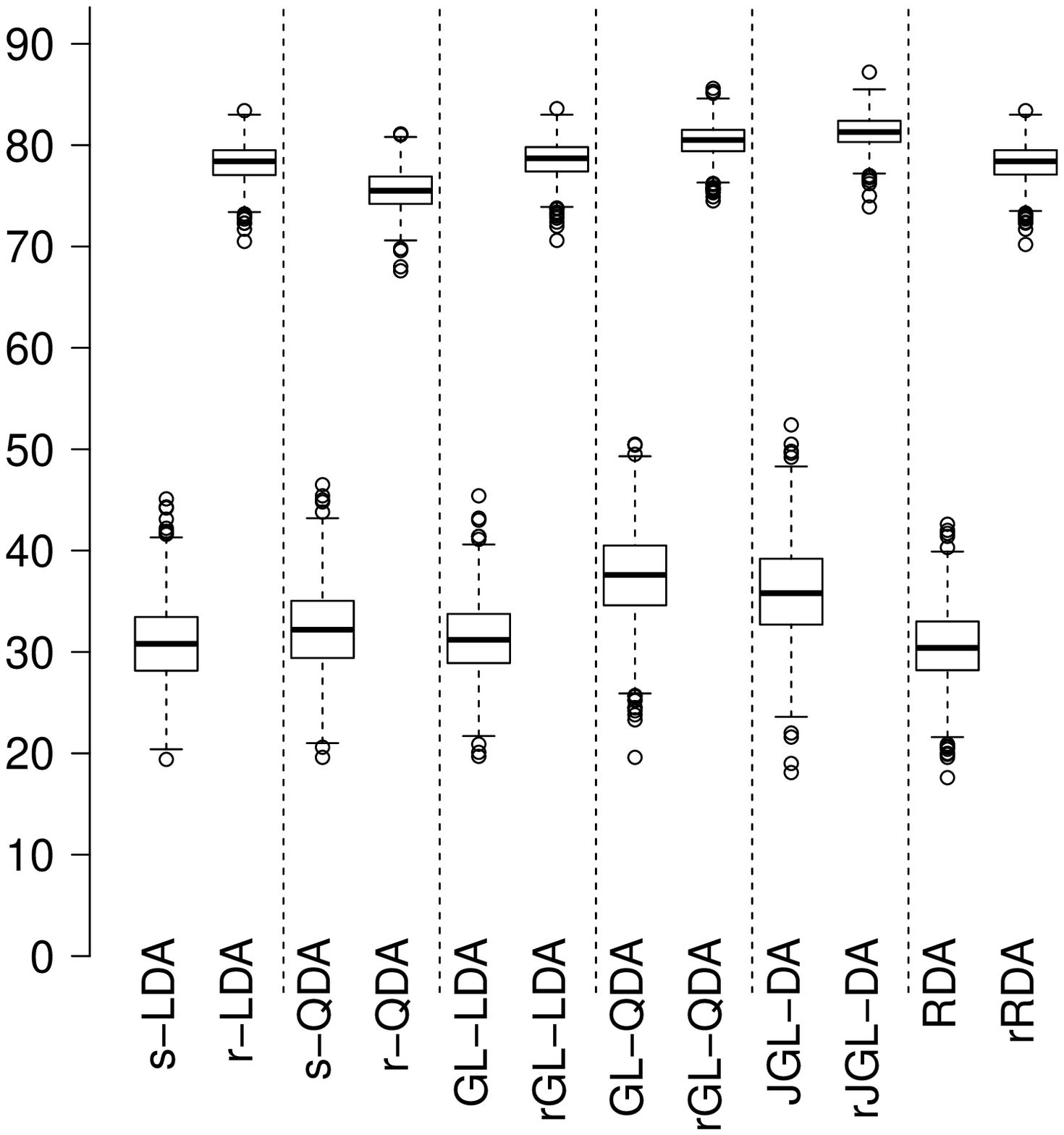}

$p=30$

\includegraphics[scale=0.33]{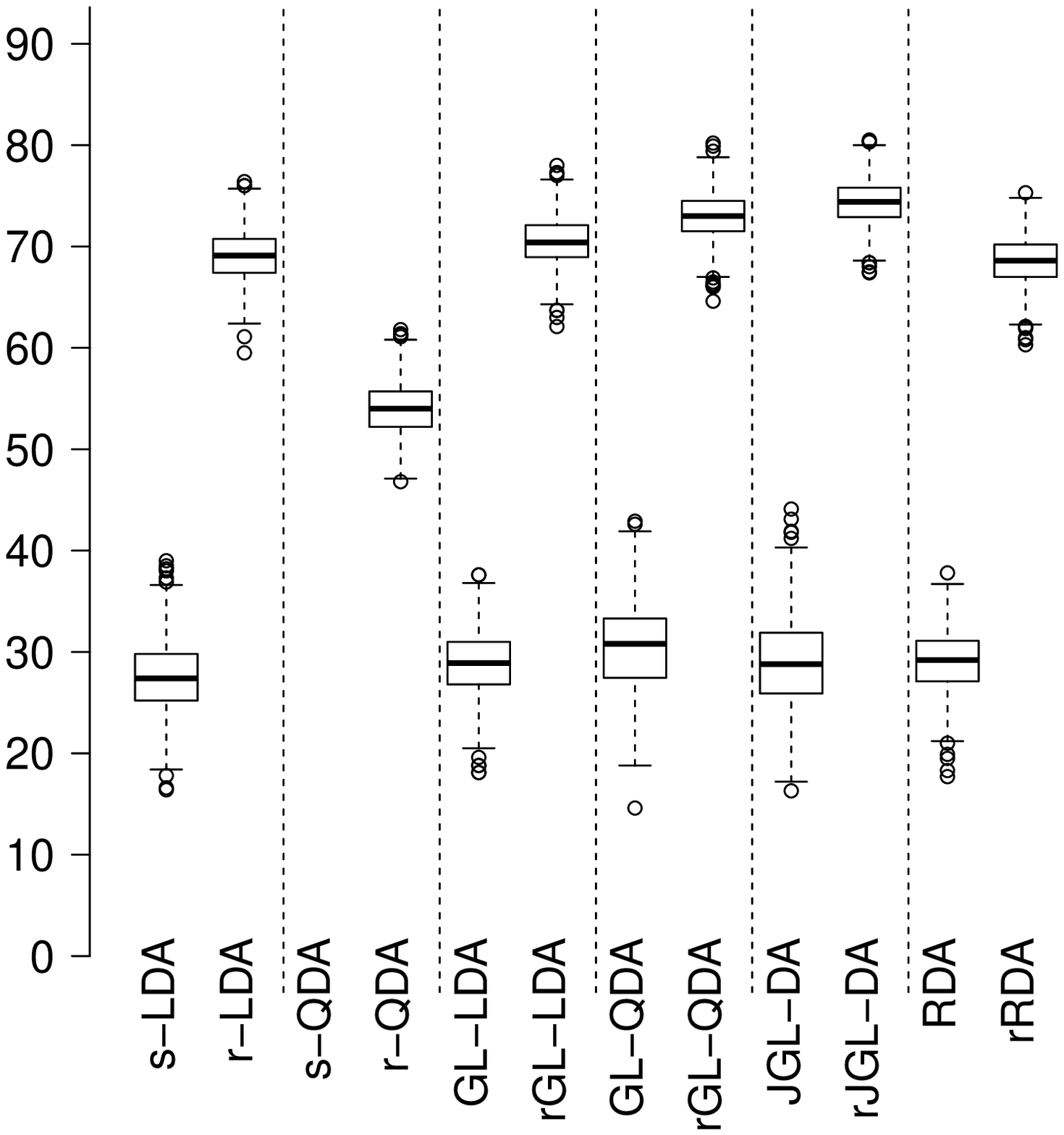}
\end{minipage}
\caption{Scenario 1 : Boxplots of correct classification percentages. Left panel : 5\% cellwise contamination, right panel: 10\% cellwise contamination.}\label{CCboxplot} 
\end{figure}

\begin{figure} 
\centering\includegraphics[scale=0.40]{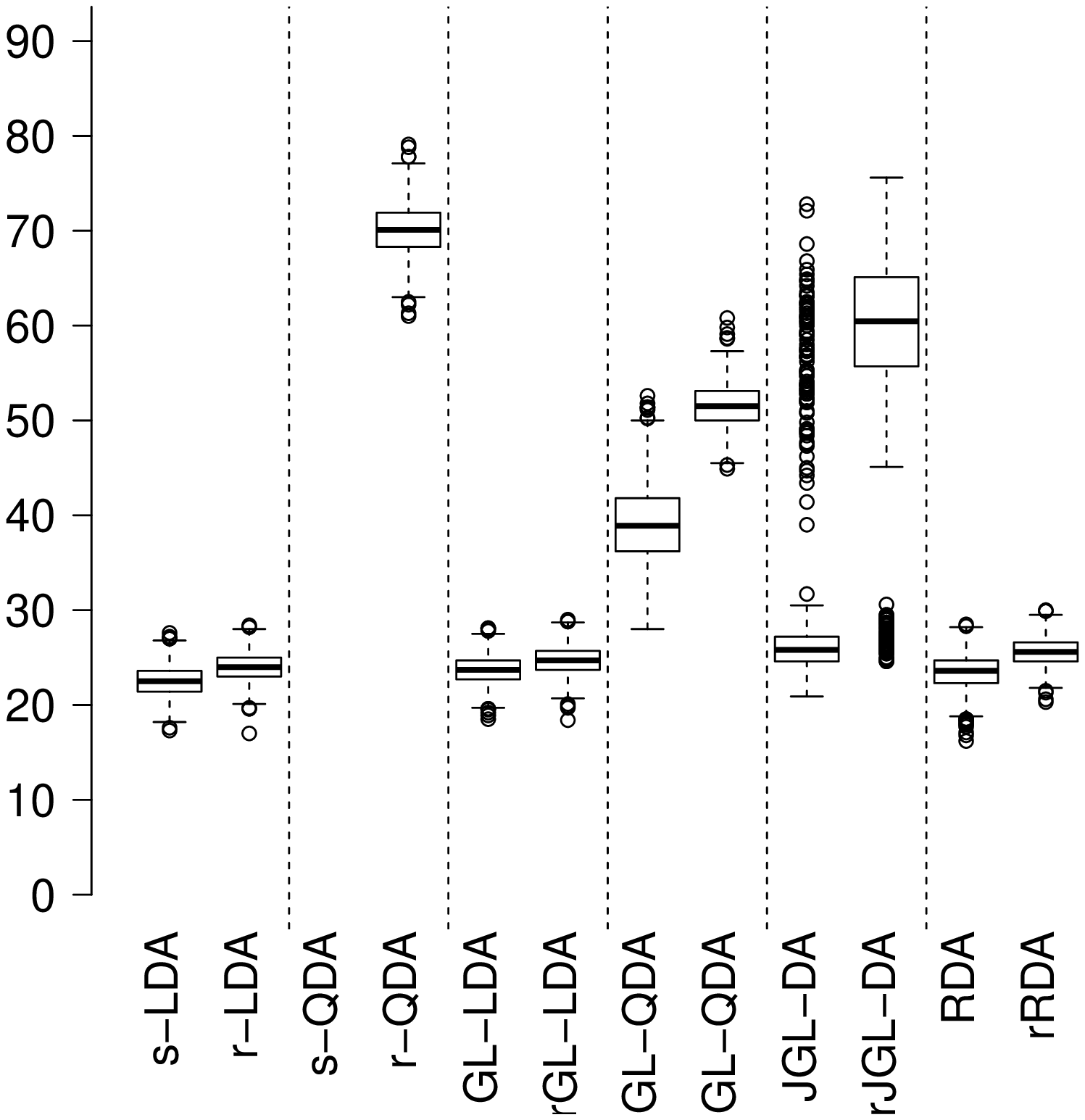}
\caption{Scenario 2 : Boxplots of correct classification percentages. 1\% cellwise contamination.}\label{CCboxplot2}
\end{figure}

\paragraph*{Results.}
We compare the non robust discriminant methods to their robust counterparts  in terms of correct classification. Figure \ref{CCboxplot} shows the results for scenario 1 (left panel: 5\% of contaminated cells, right panel: 10\% of contaminated cells), Figure \ref{CCboxplot2} for scenario 2. For each method, the boxplot of correct classification percentages of the 1000 simulation runs for the non robust version is displayed on the left while the boxplot on the right corresponds to its cellwise robust counterpart.

In all the considered contaminated settings, the cellwise robust methods maintain their good classification performance. On the contrary, the outlying cells mislead all the considered non robust methods. As a result, their correct classification percentages considerably decrease and their KL-distances (unreported) are high.

The higher the dimension $p$ and/or the higher the contamination proportion $\varepsilon$, the better the performance of the cellwise robust estimators relative to their non robust version. For instance, keeping $\varepsilon$ fixed to 5\%, rJGL-DA  leads to an increase in correct classification performance of (on average) 19 percentage points over JGL-DA in dimension $p=5$, while this gain increases to 32 percentage points in dimension $p=30$. Likewise, keeping the dimension $p=5$ fixed but varying the proportion of contaminated cells, rJGL-DA improves classification performance by 19 percentage points over JGL-DA when $\varepsilon=5\%$, and this gain doubles when $\varepsilon=10\%$.  
The deteriorating performance of the non robust methods when the dimension increases is expected in this cellwise contamination scheme. Indeed, the probability that an observation has at least one contaminated cell is $1-(1-\varepsilon)^p$. In the first scenario, for 5\% of contamination and $p=5$, already 22.6\% of the observations are expected to be contaminated, and more than 78\% if $p=30$. For 10\% of cellwise contamination, the presence of contaminated cells is expected for nearly a half of the observations if $p=5$, and almost all of them if $p=30$.

Results for the cellwise robust estimators are summarized in Table \ref{TabSimul4}. The main findings are similar to those detailed in the uncontaminated case. As the dimension increases in Scenario 1, the regularized techniques rGL-QDA and rJGL-DA  are among the best in terms of correct classification. rJGL-DA achieves considerably lower KL-distances than rGL-QDA. 
In the second scenario, r-QDA, rJGL-DA and rGL-QDA yield the best correct classification rates but the estimation accuracy of r-QDA is much worse than rJGL-DA and rGL-QDA.\\

\begin{table}[t]
\begin{center}
\begin{footnotesize}
\caption{Percentages of correct classification (CC) and KL-distance for the cellwise robust discriminant methods, averaged over 1000 simulation runs.}\label{TabSimul4}
\begin{tabular}{l|ll|cccccc}
\cline{4-9}
\multicolumn{3}{c|}{}& r-LDA & r-QDA& rGL-LDA & rGL-QDA  & rJGL-DA & rRDA\\
\hline
\multirow{6}{*}{\parbox{3cm}{Scenario 1 - $\varepsilon= 5\%$ $K=10$   }} 
& $p=5$ &CC   & 74.9 & 76.1 & 74.6 & 77.4 & 77.7 & 74.9 \\
&& KL  & 20.83 & 12.530 & 22.32 & 15.31& 14.07 & 20.95  \\  
\cline{2-9} 
&$p=10$ &CC  & 80.4 & 77.8  & 80.6 & 82.3 & 82.9& 80.4  \\ 
&& KL& 22.58 & 21.25 & 25.13 & 23.83 & 15.83 & 22.68 \\ 
\cline{2-9}
&$p=30$ & CC   & 73.5 & 56.7  & 74.9 & 76.7& 77.8& 71.6 \\
&& KL  & 33.90 & 116.03  & 40.26 & 70.25 & 23.21& 68.95  \\
    \hline
 \multirow{6}{*}{\parbox{3cm}{Scenario 1 - $\varepsilon=10\%$ $K=10$  }} 
& $p=5$ &CC & 72.0 & 73.7 & 71.7 & 75.2& 75.6& 72.0  \\  
&& KL & 26.13 & 16.41  & 27.84 & 20.47 & 18.73& 26.24  \\
\cline{2-9} 
&$p=10$ &CC  & 78.3 & 75.5  & 78.6 & 80.4 & 81.3& 78.3 \\ 
&& KL &   30.78 & 25.54  & 34.20& 33.40 & 23.38 & 30.93\\
\cline{2-9}
&$p=30$ & CC &  69.1 & 54.0  & 70.5 & 73.0 & 74.3 & 68.6\\ 
&& KL & 51.52 & 106.68  & 62.86 & 98.31 & 41.18 &85.74 \\
 \hline
 \multirow{2}{*}{\parbox{3cm}{\footnotesize Scenario 2 - $\varepsilon=1\%$ $K=6$ ~~}} 
& $p=50$ & CC & 24.0 & 70.0  & 24.7 & 51.6  & 58.6& 25.6 \\
  &&KL  & 176.60 & 227.57  & 161.81 &98.79& 117.88 & 156.82 \\ 
  \hline
\end{tabular}
    \end{footnotesize}
\end{center}
\end{table}

To summarize, the proposed cellwise robust discriminant methods have better classification performance and estimation accuracy than their non-robust counterparts in presence of cellwise outliers.  Among the robust techniques,  the regularized methods outperform the standard ones in high dimension. Among the cellwise robust regularized methods, rJGL-DA attains the best overall performance. This method relies on a similarity parameter that, when varied, covers the path from QDA to LDA. Choosing this parameter in a data-driven way makes rJGL-DA a tough competitor not only in presence of many similar precision matrices but also under LDA/QDA assumptions.

\section{Examples} \label{sec:real}

In this section, we illustrate the performance of the proposed methods on two real datasets. The first example, the \textit{forest soil} dataset, is known in the robust discriminant analysis literature. The second example, the \textit{phoneme} dataset, includes a large number of variables and has been used in papers on regularized discriminant analysis. We find most cellwise robust discriminant methods to attain better classification performance than their non robust counterparts. We also provide two visual tools showing the outlying observations and outlying cells in the analysed datasets.

\medskip

The robust mean and precision matrix estimates proposed in this article can be used for outlier detection. Usual outlier detection methods are based on the computation of distances. The sample mean and covariance matrix are, however, heavily influenced by outliers. Therefore, distances computed from them may be large for the clean observations/cells and small for the outlying ones. As a result, the actual outliers are not detected. This effect is known as the \textit{masking effect} \cite{RousseeuwLeroy}. It can be avoided by computing robust distances from robust location and precision matrix estimates like the ones proposed in Section \ref{sec:robust}.  They allow to pinpoint both outlying observations, i.e. rowwise outliers, and outlying cells, i.e. cellwise outliers. 

To find rowwise outliers in each group, robust Mahalanobis distances are computed for each observation $\bm{x}_i$ of the group,
\begin{align*}
D_{i} = \sqrt{(\bm{x}_i - \widehat{\bm{\mu}}_k)^T\widehat{\bm \Theta}_k(\bm{x}_i - \widehat{\bm{\mu}}_k) },
\end{align*}
where $\widehat{\bm{\mu}}_k$ is the vector of marginal medians in group $k$ and $\widehat{\bm \Theta}_k$ is one of the precision matrix estimates defined in Section \ref{sec:robust}. Observations with a Mahalanobis distance above the square root of the 0.99 quantile of the chi-square distribution with $p$ degrees of freedom are flagged as rowwise outliers.

To find cellwise outliers in each group, we follow the approach of \cite{Agostinelli2015}. For each cell $x_{ij}$ in group $k$, a cellwise standardized distance is computed  
\begin{align}
d_{ij} = \frac{x_{ij} - m_{k,j}}{t_{k,j}}, \label{celldist}
\end{align} 
where $m_{k,j}$ and $t_{k,j}$ estimate respectively the marginal location and scale of the $j$th variable in group $k$. We replace $m_{k,j}$ in \eqref{celldist} by the median of the $j$th variable in group $k$. For the scale, we replace $t_{k,j}$ by the square root of element $(j,j)$ of $\widehat{\bm \Theta}_k^{-1}$. Cells with standardized distance exceeding the square root of the $0.99^{1/(n_kp)}$ quantile of a chi-square distribution with one degree of freedom are flagged as cellwise outliers.

\subsection{Forest soil data}

The \textit{forest soil} dataset, available in the \texttt{R}-package \texttt{rrcovHD} \cite{rrcovHD}, contains measurements on $N=58$ soil pits in the Hubbard Brook Experimental Forest in north-central New Hampshire of 1983. For each soil sample, the exchangeable cations of  calcium, magnesium, potassium and sodium ($p=4$) are reported. The pit location can be classified in $K=3$ types of forest: spruce-fir ($n_1=11$), high elevation hardwood ($n_2=23$) and low elevation hardwood ($n_3=24$). Some unusual soil samples are present in the dataset, as already noticed in \cite{VandenBranden2005}. Note that the group sample sizes are low compared to the dimension $p$. We compare the classification performance of the proposed cellwise robust and non robust discriminant methods. The robust location and precision matrix estimates are then used to construct an outlier detection map.

\paragraph{Classification performance.} As the sample size is low, the same dataset is used to construct and evaluate the discriminant rule. Table \ref{ForestSoil} summarizes the percentages of correct classification for the non robust and robust methods. This dataset is characterized by strong overlapping groups, which causes overall low correct classification rates. We observe that the robust discriminant methods reduce the influence of the unusual soil samples and yield better correct classification rates. Although the dimension is not high, the regularized techniques rJGL-DA and rGL-QDA are to be preferred since the sample size is low.

\begin{table}
\begin{footnotesize}
\begin{center}
\caption{Percentage of correct classification. Forest soil data,  $K=3$, $p=4$, $N =58$.}\label{ForestSoil}
\renewcommand{\arraystretch}{0.8}
\begin{tabular}{llllll}
\hline
 s-LDA & s-QDA & GL-LDA & GL-QDA & JGL-DA &  RDA  \\
 56.9 & 56.9 & 56.9 & 55.1 & 56.9& 56.9 \\
\hline
\hline
r-LDA & r-QDA & rGL-LDA & rGL-QDA  & rJGL-DA &  rRDA \\
60.3 & 63.8 & 60.3 & 65.6  & 67.2& 60.3\\
\hline
\end{tabular}
\end{center}
\end{footnotesize}
\end{table}

\begin{figure}[t]
\centering\includegraphics[scale=0.7]{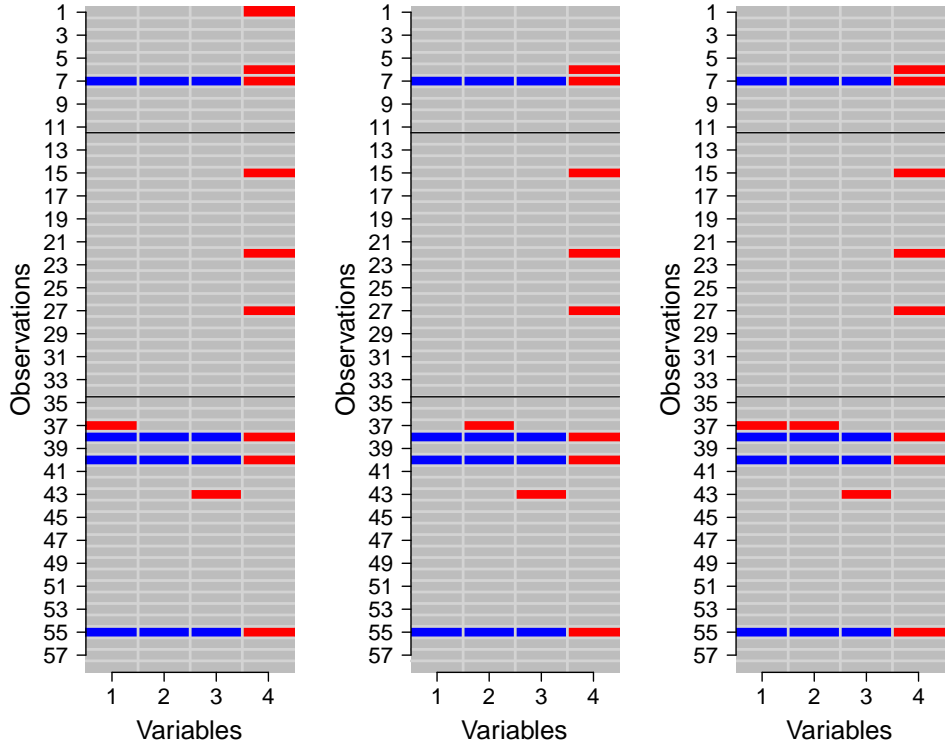}
\caption{Outlier detection map of Forest soil data: detected rowwise outliers (blue) and cellwise outliers (red) by rGL-LDA (left), rGL-QDA (middle) and rJGL-DA (right). }\label{RobDist}
\end{figure}

\paragraph*{Outlier detection.} Since the cellwise robust methods perform better, outliers might be present. To characterize the rowwise outliers, we compute robust Mahalanobis distances. To characterize the cellwise outliers, we compute robust cellwise standardized distances. Figure \ref{RobDist} visualizes the detected outliers  using the rGL-LDA (left panel), rGL-QDA (middle) and rJGL-DA (right panel) estimates. Observations flagged as rowwise outliers are colored in blue, cellwise outliers in red. The black vertical lines split the observations according to their group membership: spruce-fir (top), high elevation hardwood (middle) and low elevation hardwood (bottom).

The same four rowwise outliers are highlighted by the three estimation methods. They all have a
sodium measurement (last column) that is much higher than expected, while their other components are in line with the majority of the data. Unlike standard robust methods, the proposed cellwise discriminant methods result in less loss of information since they do not drop the entire row because of the presence of only one outlying component. The methods also detect, overall, the same outlying cells, mainly among the measurements for sodium. Used together, these rowwise and cellwise outlier detection techniques allow a deeper comprehension of the dataset.

\subsection{Phoneme data}
The \textit{phoneme} dataset \cite{Hastie2009} contains $N=1717$ observations corresponding to the record of a male voice pronouncing one of $K=2$ similar sounds, either \textit{aa} or \textit{ao}. The aim is to build a classifier of these sounds on the basis of the $p=256$ log-periodograms representing the log intensity of the sound across 256 frequencies.

\paragraph*{Classification performance.} Since there are enough observations in each group, we split the dataset into a training and a test set (with 60\%/ 40\% of the observations respectively). We evaluate the performance of the methods on the test set. To diminish the influence of the split, this procedure is repeated ten times. The average percentage of correctly classified observations is computed and reported in Table \ref{Phoneme}. For the majority of the considered methods, the robust ones perform better than their non robust counterparts. Among the cellwise robust techniques, the regularized methods rGL-LDA and rGL-QDA improve classification performance compared to respectively r-LDA and r-QDA. 

\begin{table}
\begin{footnotesize}
\begin{center}
\caption{Average percentage of correct classification.  Phoneme dataset  $K=2$, $p=256$, $N_{\rm train}=1030$, $N_{\rm test}=687$. }\label{Phoneme}
\renewcommand{\arraystretch}{0.8}
\begin{tabular}{llllll}
\hline
 s-LDA & s-QDA & GL-LDA & GL-QDA & JGL-DA & RDA \\
  77.7 & 62.4 & 81.4 & 74.9 & 78.4 & 78.2 \\
\hline
\hline
 r-LDA & r-QDA & rGL-LDA & rGL-QDA  & rJGL-DA & rRDA \\
 81.1  & 74.7  & 81.7 & 76.0 & 76.7 & 73.3
 \\
 \hline
\end{tabular}
\end{center}
\end{footnotesize}
\end{table}

\begin{figure}
\centering\includegraphics[scale=0.5]{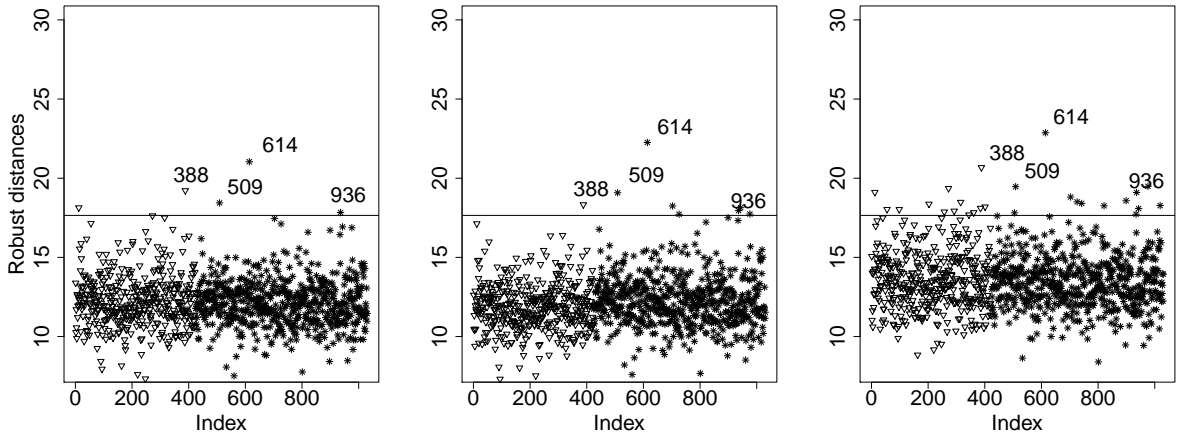}
\caption{Robust Mahalanobis distances computed with the rGL-LDA (left), the rGL-QDA (second panel) and the rJGL-DA (right panel) estimates. Observations in the first group are represented by a $\bigtriangledown$  and those in the second by a *. }\label{RobDist2}
\end{figure}

\paragraph*{Outlier detection.} As in the first data example, the robust mean and precision matrix estimates can be used to detect outliers. Here, the dataset is however too large for a clear visual inspection via the outlier detection map (cfr. Figure \ref{RobDist}). Another useful visualisation tool to detect rowwise outliers is the plot of the robust squared Mahalanobis distances versus the observation numbers, as represented in Figure \ref{RobDist2}. The distances are computed with the rGL-LDA estimates in the left panel, and with the rGL-QDA and rJGL-DA estimates in the middle and right panels. The vertical line corresponds to the 0.99 chi-square quantile with $p=256$ degrees of freedom. Observations beyond this threshold are considered as rowwise outliers. For all the methods, outliers are detected in both groups. The same extreme outlying rows are highlighted by all the estimation methods.

The different methods also detect overall the same  outlying cells (not shown), although JGL-DA highlights many more cellwise outliers. Again, the multivariate outlying behaviour of some entire rows may be explained by only one abnormal component, as is the case for observations 388, 509 and 614.  On the contrary, the extreme rowwise outlier 936  has a high Mahalanobis distance while none of its components is flagged by the cellwise outlier detection procedure. Its outlying behaviour is thus caused by a particular relation between the components rather than by one of its components. Hence, it is important to consider both the rowwise and cellwise outlier detection procedures in combination.

\section{Discussion}\label{sec:ccl}

In this paper, cellwise robust discriminant analysis methods are proposed. We discuss all the implementation issues and we make the code publicly available. The proposed discriminant methods enjoy several important advantages. They are robust against cellwise outliers, a type of outliers that is very likely to occur in high-dimensional datasets. We provide visual tools to detect both rowwise and cellwise outliers (cfr. Section \ref{sec:simul}). Furthermore, contrary to LDA that makes the strong assumption of homoscedasticity, we consider methods that emphasizes the true difference between the group covariance matrices while making use of their similarities. As such, the proposed methods lie in between LDA and QDA. Our simulations show that these approaches result in better classification performance as well as higher estimation accuracy. 
Finally, many discriminant methods require dimension reduction techniques to be computable in high-dimension (see \cite{HubertEngelen}, \cite{VandenBranden2005}). Our methods, in contrast, are computable even when the dimension exceeds the group sample size without requiring any preprocessing step. By using sparse precision matrix estimates, we reduce the effect of uninformative variables. 
	
The proposed regularized methods depend on regularization parameters that are tuned in a data-driven way. In this paper, we focus on parameter selection via minimization of the BIC. In the classification context, one may also be interested in selecting parameters so as to maximize the expected out-of-sample correct classification rate, which can be obtained by $L$-fold cross validation. This method is, however, much more time consuming. Furthermore, in our simulations, it, overall, does not outperform BIC model selection in terms of correct classification. Even more, parameter selection by $L$-fold cross validation sometimes achieve higher KL-distances. 

\paragraph*{Acknowledgements.}

We thank Prof. C. Croux who provided insights and critical comments that helped improving the manuscript. We gratefully acknowledge support from the FWO (Research Foundation Flanders, contract number 12M8217N).

\bibliographystyle{plainnat}
\bibliography{bibJGLDA}

\begin{thebibliography}{27}
\providecommand{\natexlab}[1]{#1}
\providecommand{\url}[1]{\texttt{#1}}
\expandafter\ifx\csname urlstyle\endcsname\relax
  \providecommand{\doi}[1]{doi: #1}\else
  \providecommand{\doi}{doi: \begingroup \urlstyle{rm}\Url}\fi

\bibitem[Agostinelli et~al.(2015)Agostinelli, Leung, Yohai, and
  Zamar]{Agostinelli2015}
C.~Agostinelli, A.~Leung, C.J. Yohai, and R.~H. Zamar.
\newblock Robust estimation of multivariate location and scatter in the
  presence of cellwise and casewise contamination.
\newblock \emph{TEST}, 24\penalty0 (3):\penalty0 441--461, 2015.

\bibitem[Alqallaf et~al.(2002)Alqallaf, Komis, and Zamar]{Alqallaf2002}
F.~A. Alqallaf, Martin~R.D. Komis, K.~P., and R.H. Zamar.
\newblock Scalable robust covariance and correlation estimates for data mining.
\newblock \emph{Proceedings of the Eighth ACM SIGKDD International Conference
  on Knowledge Discovery and Data Mining}, pages 14--23, 2002.

\bibitem[Alqallaf et~al.(2009)Alqallaf, Van~Aelst, Yohai, and
  Zamar]{Alqallaf2009}
F.A. Alqallaf, S.~Van~Aelst, V.J. Yohai, and R.H. Zamar.
\newblock Propagation of outliers in multivariate data.
\newblock \emph{The Annals of Statistics}, 37:\penalty0 311--331, 2009.

\bibitem[Croux and Dehon(2001)]{CrouxDehon}
C.~Croux and C.~Dehon.
\newblock Robust linear discirminant analysis using {S}-estimators.
\newblock \emph{The Canadian Journal of Statistics}, 29\penalty0 (3):\penalty0
  473--493, 2001.

\bibitem[Croux and Öllerer(2015)]{Ollerer2015}
C.~Croux and V.~Öllerer.
\newblock \emph{Modern Multivariate and Robust Methods}, chapter Robust
  high-dimensional precision matrix estimation.
\newblock Springer, 2015.

\bibitem[Danaher(2013)]{JGL}
P.~Danaher.
\newblock \emph{JGL: Performs the Joint Graphical Lasso for sparse inverse
  covariance estimation on multiple classes}, 2013.
\newblock URL \url{https://CRAN.R-project.org/package=JGL}.
\newblock R package version 2.3.

\bibitem[Danaher et~al.(2014)Danaher, Wang, and Witten]{Danaher2014}
P.~Danaher, P.~Wang, and D.~Witten.
\newblock The joint graphical lasso for inverse covariance estimation across
  multiple classes.
\newblock \emph{Journal of the Royal Statistical Society, Series B},
  76:\penalty0 373--397, 2014.

\bibitem[Filzmoser and Fritz(2006)]{pcaPP}
P.~Filzmoser and H.~Fritz.
\newblock \emph{pcaPP : Robust {PCA} by Projection Pursuit}, 2006.
\newblock URL \url{https://CRAN.R-project.org/package=pcaPP}.
\newblock R package version 1.0.

\bibitem[Filzmoser et~al.(2008)Filzmoser, Maronna, and Werner]{Filzmoser2008}
P.~Filzmoser, R.~Maronna, and M.~Werner.
\newblock Outlier identification in high dimension.
\newblock \emph{Computational Statistics and Data Analysis}, 52:\penalty0
  1694--1711, 2008.

\bibitem[Filzmoser et~al.(2012)Filzmoser, Hron, and Templ]{Filzmoser}
P.~Filzmoser, K.~Hron, and M.~Templ.
\newblock Discriminant analysis for compositional data and robust parameter
  estimation.
\newblock \emph{Computational Statistics}, 27:\penalty0 585--604, 2012.

\bibitem[Friedman et~al.(2008)Friedman, Hastie, and Tibshirani]{Friedman2008}
J.~Friedman, T.~Hastie, and R.~Tibshirani.
\newblock Sparse inverse covariance estimation with the graphical lasso.
\newblock \emph{Biostatistics}, 9\penalty0 (3):\penalty0 432--441, 2008.

\bibitem[Friedman(1989)]{Friedman1989}
J.~H. Friedman.
\newblock Regularized discriminant analysis.
\newblock \emph{Journal of the American Statistical Association}, 84:\penalty0
  165--175, 1989.

\bibitem[Gao et~al.(2016)Gao, Zhu, She, and Pan]{Gao2016}
C.~Gao, Y.~Zhu, X.~She, and W.~Pan.
\newblock Estimation of multiple networks in gaussian mixture models.
\newblock \emph{Electronic Journal of Statistics}, 10:\penalty0 1133--1154,
  2016.

\bibitem[Hastie et~al.(2009)Hastie, Tishirani, and Friedman]{Hastie2009}
T.~Hastie, R.~Tishirani, and J.~Friedman.
\newblock \emph{The Elements of Statistical Learning, Data Mining, Inference
  and Prediction, Second Edition}.
\newblock Springer Verlag, New York, 2009.

\bibitem[Hubert and Engelen(2004)]{HubertEngelen}
M.~Hubert and S.~Engelen.
\newblock Robust {PCA} and classification in biosciences.
\newblock \emph{Bioinformatics}, 20\penalty0 (11):\penalty0 1728--1736, 2004.

\bibitem[Hubert and Van~Driessen(2004)]{HubertVanDriessen}
M.~Hubert and K.~Van~Driessen.
\newblock Fast and robust discriminant analysis.
\newblock \emph{Computational Statistics and Data Analysis}, 45\penalty0
  (2):\penalty0 301--320, 2004.

\bibitem[Price et~al.(2015)Price, Geyer, and Rothman]{Price2015}
B.~Price, C.~Geyer, and A.~Rothman.
\newblock Ridge fusion in statistical learning.
\newblock \emph{Journal of Computational and Graphical Statistics}, 24\penalty0
  (2):\penalty0 439--454, 2015.

\bibitem[Rousseeuw and Croux(1993)]{RousseeuwCroux1993}
P.~Rousseeuw and C.~Croux.
\newblock Alternatives to the median absolute deviation.
\newblock \emph{Journal of the American Statistical Association}, 88\penalty0
  (424):\penalty0 1273--1283, 1993.

\bibitem[Rousseeuw and Leroy(1987)]{RousseeuwLeroy}
P.~J. Rousseeuw and A.M. Leroy.
\newblock \emph{Robust regression and outlier detection}.
\newblock John Wiley and Sons, New-York, 1987.

\bibitem[Tarr et~al.(2015)Tarr, Müller, and Weber]{Tarr2015}
G.~Tarr, S.~Müller, and N.C. Weber.
\newblock Robust estimation of precision matrices under cellwise contamination.
\newblock \emph{Computational Statistics and Data Analysis}, 93:\penalty0
  404--420, 2015.

\bibitem[Todorov(2016)]{rrcovHD}
V.~Todorov.
\newblock \emph{rrcovHD : Robust Multivariate Methods for High dimensional
  data}, 2016.
\newblock URL \url{https://CRAN.R-project.org/package=rrcovHD}.
\newblock R package version 0.2-5.

\bibitem[Van~Aelst(2016)]{VanAelst2015}
S.~Van~Aelst.
\newblock Stahel-{D}onoho estimation for high dimensional data.
\newblock \emph{International Journal of Computer Mathematics}, 93:\penalty0
  628--639, 2016.

\bibitem[Vanden~Branden and Hubert(2005)]{VandenBranden2005}
K.~Vanden~Branden and M.~Hubert.
\newblock Robust classification in high dimension based on the {SIMCA} method.
\newblock \emph{Chemometrics and Intelligent Laboratory Systems}, 79:\penalty0
  10--21, 2005.

\bibitem[Xu et~al.(2014)Xu, Huang, I., Liu, Sun, and Satoshi]{Xu2014}
B.~Xu, K.~Huang, King I., C.~Liu, J.~Sun, and N.~Satoshi.
\newblock Graphical lasso quadratic discriminant function and its application
  to character recognition.
\newblock \emph{Neurocomputing}, 129:\penalty0 33--40, 2014.

\bibitem[Yuan and Wang(2013)]{YuanWang2013}
T.~Yuan and J.~Wang.
\newblock A coordinate descent algorithm for sparse positive definite matrix
  estimation.
\newblock \emph{Statistical Analysis and Data Mining}, 6\penalty0 (5):\penalty0
  431--442, 2013.

\bibitem[Zhao et~al.(2012)Zhao, Liu, Roeder, Lafferty, and Wasserman]{Zhao2012}
T.~Zhao, H.~Liu, K.~Roeder, J.~Lafferty, and L.~Wasserman.
\newblock The huge package for high-dimensional undirected graph estimation in
  {R}.
\newblock \emph{Journal of Machine Learning Research}, 13:\penalty0 1059--1062,
  2012.

\bibitem[Zimek et~al.(2012)Zimek, Schubert, and Kriegel]{Zimek2012}
A.~Zimek, E.~Schubert, and H-P. Kriegel.
\newblock A survey on unsupervised outlier detection in high dimensional
  numerical data.
\newblock \emph{Statistical Analaysis and Data Mining}, 5:\penalty0 363--476,
  2012.

\end{thebibliography}

\end{document}